\documentclass{iopjournal}

\usepackage{orcidlink}
\usepackage{cite}
\usepackage{lmodern}
\usepackage{graphicx} 	
\usepackage{iftex}
\usepackage{color} 		
\usepackage{url}		
\usepackage{extarrows} 	
\usepackage{framed} 	
\usepackage{chemfig} 

\usepackage{soul}
\usepackage{mathrsfs}

\usepackage{siunitx}

\usepackage{dcolumn}
\usepackage{booktabs}
\usepackage{adjustbox}
\usepackage{multirow}
\usepackage{indentfirst}
\usepackage{ragged2e}

\usepackage{bm}       
\usepackage{amssymb,amsmath,nccmath} %
\usepackage{amsthm}   
\usepackage{amsfonts} %
\usepackage{mathrsfs} %
\usepackage{esint} 

\usepackage{amscd}    
\usepackage{braket}
\usepackage{slashed}
\usepackage{bbm}
\usepackage{bropd}
\usepackage{float}             

\usepackage{mathtools,tikz}
\usetikzlibrary{patterns}
\usepackage{pgf}
\usepackage{pgfplots}
\pgfplotsset{compat=1.18}
\usetikzlibrary{positioning,fit}

\begin{document}

\articletype{Paper}

\title{Spatial distribution of Majorana modes in extended-range Kitaev finite chains}

\author{Pedro B. Widniczck$^{1, 2}$ 
\orcidlink{0009-0006-0834-2816}, 
Gerardo Mart\'{\i}nez$^{1}$ 
\orcidlink{0000-0002-3871-7598}}




\affil{$^1$Instituto de F\'{\i}sica, Universidade Federal do Rio Grande do Sul, Porto Alegre RS, 91501-970, Brasil}

\affil{$^2$Institute for Theoretical Physics, Utrecht University, 
3584 CC Utrecht, The Netherlands}
\date{\today}

\begin{abstract}
    We investigate the topological properties of finite Kitaev chains with extended-range interactions, focusing on regimes where the topological winding number is preserved. We present an explicit method to evaluate this topological invariant and determine the quantum phase transitions across several representative cases. Using the local density of states and Majorana polarization, we characterize the average spatial distribution of zero-energy modes and connect it to both the ground-state fermion parity and edge-mode occupation numbers through an effective parity model. This effective parity undergoes energetic and functional changes across ground-state fermion parity switches. For finite next-nearest-neighbor Kitaev chains, we compute zero-energy modes in regimes of physical interest and find a direct correlation between ground-state fermion parity, edge occupation numbers, and quantum phase transitions with central charge $c=1$, appearing as a region of local extrema in the average edge-mode distribution. Finally, we show that single-mode localization fails to describe ground-state physics when multiple near-degenerate Majorana modes arise.
\end{abstract}

\section{Introduction}

The topological behavior seen in condensed matter systems has become increasingly important, attracting growing attention in recent decades, motivated by technological advances. 
Many quantum topological materials and their emerging phenomena have been described in terms of a range of quantum Hall effects~\cite{vonKlitzing1986TheEffect, Stormer1999TheEffect, Bernevig2006QuantumWells, Haldane1988ModelAnomaly}. 
The exotic properties of excitations in topological insulators and topological superconductors are understood, with the Cartan symmetry classification~\cite{Kitaev2009PeriodicSuperconductors, Ryu2010TopologicalHierarchy, Bernevig2013TopologicalSuperconductors}, linking  dimensionality and symmetry classes of non-interacting systems with the nature of topological invariants, generalizing symmetry protected topological (SPT) phases. 
For superconducting systems, Kitaev introduced, in a seminal work~\cite{Kitaev2001UnpairedWires}, one of the simplest topological models and provided a natural interpretation of topological superconductivity: the existence of unpaired Majorana edge modes. 
In that work, spinless electrons in one dimension (1D) are coupled with hopping and pairing between nearest neighbors, yielding a topological phase transition without a local order parameter. 

The natural framework for understanding superconducting systems that exhibit topological behavior is via Majorana operators. 
These operators behave as fermions, and they must pair up to form the usual excitations in metallic or insulating systems, similar to the mixing of creation and annihilation operators in the Bogoliubov formalism for normal superconductors. 
They provide a useful scheme to describe new emerging phenomena in topological systems exhibiting unpaired Majorana modes, or more generally Majorana zero modes (MZM)~\cite{Beenakker2013SearchSuperconductors, Leijnse2012IntroductionFermions, Fidkowski2012UniversalJunctions}. 
Such states are not only of fundamental interest, they also provide a promising platform for fault-tolerant quantum computing~\cite{tenHaaf2025ObservationChain, Aghaee2025InterferometricDevices_short, Sarma2015MajoranaComputation, Bravyi2002FermionicComputation, Aghaee2023InAs-AlProtocol_short}. 
Subsequently, new ingredients were added to Kitaev chains, such as higher dimensional variations, or including spin-orbit interactions, or extended cluster spin systems via a Jordan-Wigner transformation, or coupled Kitaev chains (Kitaev ladders) or other general extensions of Majorana wires~\cite{Sigrist1991PhenomenologicalSuperconductivity, Wakatsuki2014FermionSuperconductor, Sedlmayr2015VisualizingPolarization, Bena2017TestingPolarization, Ezawa2017ExactSuperconductors, Maiellaro2018TopologicalLadder, Nehra2019EnhancementLeads, Verresen2018TopologyChains, Jones2019AsymptoticSystems, Jones2023Bulk-BoundaryChains, Jones2023ExactChains, Kanehira2023JosephsonSuperconductors, Roy2023CriticalPotential, Choi2024Finite-temperatureDimension, Fidkowski2011TopologicalDimension, Alecce2017ExtendedPairing, Li2021TopologicalCouplings,  Ghuneim2026ZeroChain, Pathak2026AnalyticalInward, DeGottardi2013MajoranaLandscapes}, whose main aims are to produce more realistic set-ups to experimentally assess the topological information encoded in the ground state of many-body materials. 

In spite of these theoretical advances, experimental validation remains a major challenge due to the fragility of MZM and noise present in the experimental setups; moreover, other boundary effects such as Andreev bound states~\cite{Sato2011TopologyDispersion, Fidkowski2012UniversalJunctions, Sato2017TopologicalReview, Mondal2025DistinguishingHeterostructure, tenHaaf2025ObservationChain, Aghaee2025InterferometricDevices_short, DeGottardi2013MajoranaLandscapes, Grabsch2019PfaffianDetection, Roy2026QuantumNanowires} display significant uncertainty for definite validation of MZM. 
Nowadays, the most clear-cut technique for probing topological edge states uses scanning tunneling microscopy~\cite{Balatsky2006Impurity-inducedSuperconductors, Fischer2007ScanningSuperconductors}, which obtains the spatial distribution of electronic states using the tunneling current between the sample and the tip in a range of voltage bias: detecting a peak at zero-bias in localized regions suggests MZM presence in topological superconductors. 

Despite all these variants of the Kitaev chain, the general spatial behavior of Majorana modes in finite chains is frequently unaddressed. 
This perspective is of particular interest in the context of physical applications, since the overlap of the Majorana modes is directly related to the ground-state fermion parity and possibly to its definite measurement, see~\cite{Fidkowski2012UniversalJunctions, tenHaaf2025ObservationChain, Aghaee2025InterferometricDevices_short, Grabsch2019PfaffianDetection, Roy2026QuantumNanowires}.

We summarize our results as follows.
Given the existence of multiple topological phases in long-range Kitaev systems, we introduce a method to compute the topological invariant, thereby construct the phase diagrams for several scenarios by computing the winding number and the central charge at the phase transitions.  
Analyzing the Majorana average position and the effective parity model in the original Kitaev chain, we re-obtain known results of the correspondence between the Kitaev chain and the XY spin model in 1D. 
Including next-nearest-neighbor interactions, we find a physical consequence derivable from the topological invariant: changing  topological phases with opposite index values induces a phase factor in the wavefunction that swap the Majorana modes from one edge to the other.
On the other hand, in phases with multiple Majorana modes, we encounter regions containing many such modes, where the nearly degenerate eigenmodes become intertwined. 
Consequently, single-particle properties (as characterized by the Majorana average position and Majorana polarization) fail to capture the interference between them; this interference, however, can be observed through the local density of states. 

The paper is organized as follows.
Section~\ref{sec: ERKC} defines the class of extended-range models and topological quantities we study, and compute the phase diagrams in several scenarios. 
Section~\ref{sec: Maj-form} introduces the Majorana basis formalism, identifies the numerical diagonalization procedure, and defines the edge modes and ground-state properties of interest. 
Finally, Section~\ref{sec: finiteC} discusses the properties of finite chains using the Majorana framework to describe the topological properties of the Majorana modes in finite chains for the scenarios addressed in Section~\ref{sec: ERKC}.

\section{Extended-range Kitaev chain} 
\label{sec: ERKC}

Consider spinless fermions in a 1D lattice with $L$ sites whose hopping and pairing extend to the $q$ nearest neighbors~\cite{Fidkowski2011TopologicalDimension, DeGottardi2013MajoranaLandscapes, Alecce2017ExtendedPairing, Li2021TopologicalCouplings, Ghuneim2026ZeroChain, Pathak2026AnalyticalInward}.
Its Hamiltonian reads 

\begin{align}
    H &= - \mu \sum_{j = 1}^L c_j^\dagger c_j^{\,} - \sum_{j = 1}^{L-1} \sum_{\ell=1}^{q} \big( t_\ell^{} c_j^\dagger c_{j+\ell}^{\,} - \Delta_\ell^{} c_{j}^\dagger c_{j+\ell}^\dagger + \mbox{H.c.} \big), \label{eq: Genham KC}
\end{align}
with algebraically decaying couplings for the pairing and the hopping, respectively, given by
    
\begin{align}
    \Delta_\ell &= \Delta \left(\frac{1}{\ell}\right)^\alpha\,, &
    t_\ell &= t \left(\frac{1}{\ell}\right)^\beta\,.
    \qquad\qquad\qquad\qquad
    \label{eq: EKM}
\end{align}
\noindent
The exponents $\alpha$ and $\beta$ are free parameters that can vary within the interval $[0,\infty)$.

\subsection{Topological invariant}

The topological properties of such a family of Hamiltonians, in the thermodynamic limit ($L\rightarrow\infty$), are easily computed in momentum space. 
We express the Hamiltonian as $H = \sum_k \widehat{\psi}_k^{\,\dagger}\, h_k^{} \,\widehat{\psi}_k^{} = \sum_k \widehat{\psi}_k^{\,\dagger}  (\mathbf{n}_k^q \cdot 
    \boldsymbol{\tau}) \, \widehat{\psi}_k^{}$, with

\begin{align}
    h_k &= 
    - \bigg[ \frac{\mu}{2} + \sum_{\ell=1}^q t_\ell \cos{(\ell k)} \bigg] \tau_z + \sum_{\ell=1}^q  \Delta_\ell \sin{(\ell k)}\, \tau_y. \label{eq: ham-windingvec}  
\end{align}
where $k = 2\pi n/L$ is the momentum within the Brillouin Zone (BZ) and $\boldsymbol{\tau}$ is the vector of Pauli matrices. 
The norm $|\mathbf{n}_k^q|$ expresses the eigenenergies of the bands, and its angle $\varphi_k^q$ provides the eigenstates and, consequently, its topological character. 
Recall that the system topology is usually not obtained from the eigenenergies (although it is sometimes possible~\cite{Bernevig2013TopologicalSuperconductors, Cayssol2021TopologicalTheory}). 
The topological angle

\begin{align}
    \varphi_k^q &= -\tan^{-1} \bigg[ \frac{\Delta/t \sum_{\ell=1}^q \sin{(\ell k)}/\ell^{\alpha} }{\mu/(2t) + \sum_{\ell=1}^{q} \cos{(\ell k)}/\ell^{\beta} } \bigg], 
    \label{eq: varphi-Gen}
\end{align}
provides the components $\phi_{kl}$ and $\psi_{kl}$ of the eigenstates in Periodic Boundary Conditions, defined in Eq.~(\ref{eq: Bogo-excitations}). Two-band systems of the BDI symmetry class have their topological invariants described by the winding number~\cite{Nakahara1990GeometryPhysics}, defined as (we will shorthand $\varphi_k^q$ as $\varphi_k$)

\begin{align}
    \nu &\equiv \frac{1}{2\pi} \int_{BZ} d \varphi_k .
    \label{topo_invariant}
\end{align}
This invariant counts the number of revolutions of $\mathbf{n}_k$ around the BZ origin. 
We calculate the topological index $\nu$ by looking for discontinuities in the angle $\varphi_k$ within the BZ. 
Integrating $d\varphi_k/dk$ may allow $m$ pairs of discontinuities.  
The topological index is then obtained by analyzing the one-sided limits $\lim_{\epsilon \rightarrow 0^+} \varphi_{l_i \pm \epsilon} \equiv \varphi_{l_i,\pm}$, where we define $+$ as the limit on the right side and $-$ as the limit on the left side of the discontinuities. 
The topological invariant is then calculated using the expression

\begin{align}
    \nu &= \frac{1}{2\pi} \sum_i \int_{l_{i}}^{l_{i+1}} \od{\varphi_k}{k} dk = \frac{1}{2\pi} \sum_{i=0}^{2m} \br{\varphi_{l_{i+1},-} - \varphi_{l_i,+}}, \label{eq: disc nu}
\end{align}
where $l_i$ are the points (ordered) in the BZ with $n_x = 0$.

\subsection{Critical topological states}

Beyond the gapped phases, critical points might also exhibit different topological behaviors in terms of gapless SPT phases~\cite{Verresen2018TopologyChains, Jones2019AsymptoticSystems, Jones2023Bulk-BoundaryChains, Jones2023ExactChains}. 
This implies that topological phase transitions between distinct topological phases cannot be deformed into each other~\cite{Verresen2018TopologyChains}. 
The character of the phase transition is described by the holomorphic function 

\begin{align}
    f(z) &= \sum_{a} M_{a} z^a, \label{eq: holo-func}
\end{align}
where $M_a = M_{i,i+a}$ is determined from the matrix elements given in Eq.~(\ref{eq: Ham_mat}) below. 
This approach allows the introduction of a topological invariant in gapless systems, which is given by the central charge $c = N_u/2$, where $N_u$ is the number of zeros of $f(z)$ at the unit circle, obtained by considering parameterization $z = e^{ik}$. 
The formalism also implies that $\omega = N_z - q$, where $N_z$ describes the number of zeros of $f(z)=0$ within the unit circle, is also a topological invariant\footnote{Of course, the number of zeros inside the unit circle cannot change, unless we cross the unit circle (gapless point). 
Moreover, the closer value $z_0$ to the unit circle gives the correlation length~\cite{Jones2019AsymptoticSystems, Verresen2018TopologyChains,  Jones2023Bulk-BoundaryChains, Jones2023ExactChains}. }. 
The central charge is different from zero at critical points (quantum phase transitions)~\cite{PasqualeCalabrese2004EntanglementTheory, Verresen2018TopologyChains}, and $\omega$ is equivalent to the winding number $\nu$, Eq.~(\ref{eq: disc nu}), for gapped phases. Contrary to the winding number ($\nu$), $\omega$ is defined at critical points, thus despite being equivalent at gapped phases, we keep it as a separate topological index.

\subsection{Phase diagrams}

The formalism presented above is valid for 1D two-band systems in the BDI symmetry class. 
The topological properties are described from Eq.~(\ref{eq: varphi-Gen}) and Eq.~(\ref{eq: holo-func}). 
In the gapped phases, the condition $\cos{\varphi_{k_i}}=0$ identifies the discontinuities in the topological angle $\varphi_{k_i}$, while the analysis of $\sin{\varphi_{k_i}}=0$ accounts for a sign factor in terms of the sum defining the topological invariant, Eq.~(\ref{topo_invariant}). 
For more details, see App.~\ref{ap: trunc-q}.

\begin{figure}[!hbt]
    \centering
    \includegraphics[width= 0.8\linewidth]{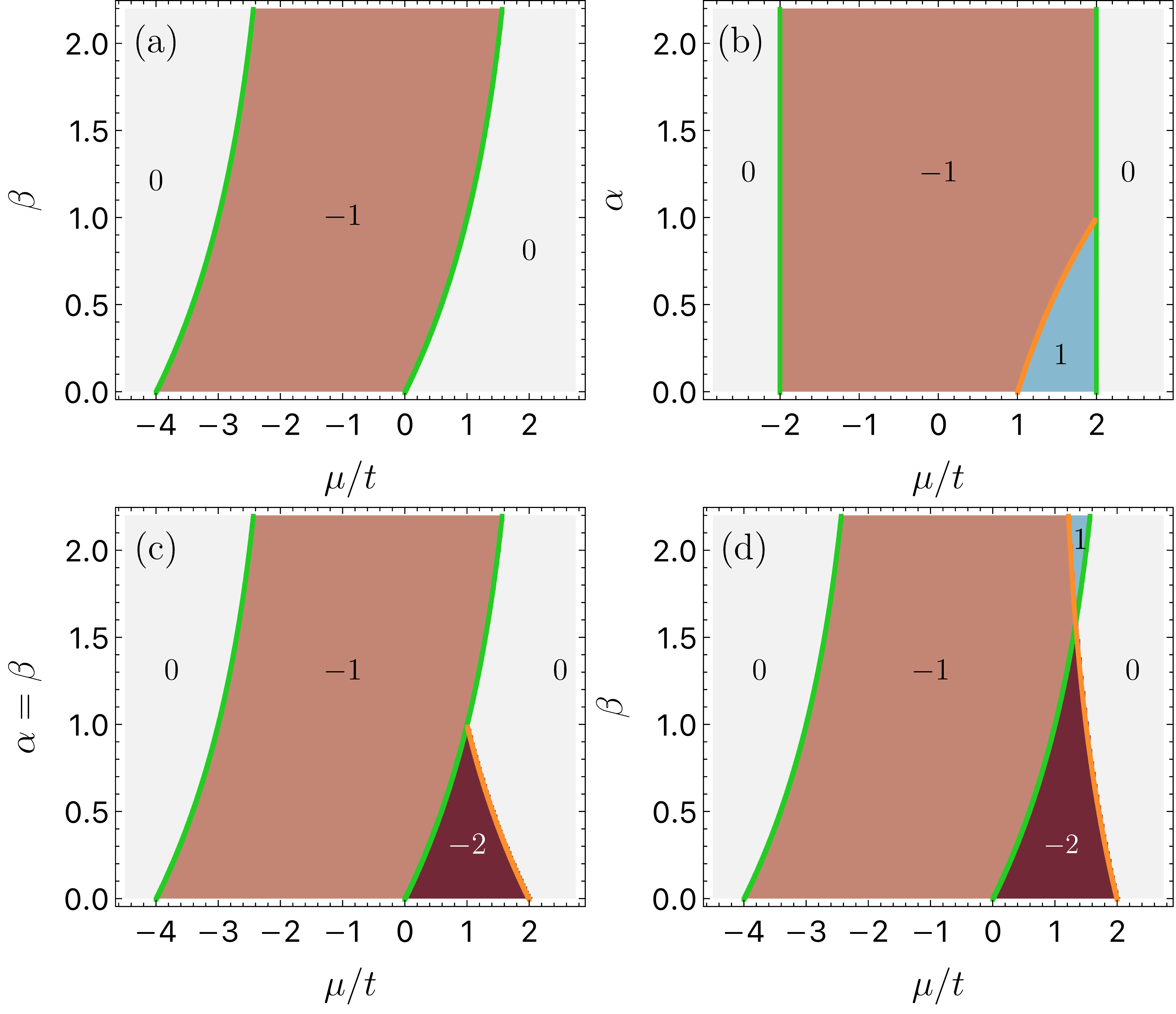}
    \caption{\justifying Topological phase diagrams of the extended Kitaev chain with next-nearest neighbors $(q=2)$; numbers indicate different topological indices:
    (a) $\alpha \rightarrow \infty$, for running $\beta$.  
    (b) $\beta \rightarrow \infty$, for running $\alpha$. 
    (c) The homogeneous chain $\alpha=\beta$. 
    (d) $\alpha = 0$ and running $\beta$. Panel (a) is similar to the original Kitaev chain, yet with bent critical Ising lines. Green curves separate different topological phases with $|\Delta\nu|= 1$ and exhibit a central charge $c=1/2$. Panels~(b) and~(d) show topological phases with chiral behavior, $\nu=\pm 1$, while panels~(c) and~(d) exhibit new topological phases with $\nu = -2$. New kind of topological transition lines are depicted as orange curves separating topological phases with $|\Delta\nu|= 2$, which exhibit a central charge $c=1$. 
    }
    \label{fig: Diagrams}
\end{figure}

Consider the case $q=2$ to be valid, Fig.~\ref{fig: Diagrams} exhibit the distinct phase diagrams in several regimes of the next-nearest-neighbors hopping and pairing terms, controlled by the exponents $\alpha$ and $\beta$ defined in Eq.~(\ref{eq: EKM}). 
We observe that different topological phases arise when $\alpha<1$ or $\beta<1$. 
In the limit $\alpha \rightarrow \infty$, Fig.~\ref{fig: Diagrams}(a) displays the phase diagram showing the usual topological phase with $\nu=-1$; now the critical points, separating distinct topological phases, are functions of the exponent $\mu_c = \mu_c(\beta)$, as the green lines show. 
When $\beta \rightarrow \infty$, Fig.~\ref{fig: Diagrams}(b) exhibit changes in the sign of the topological invariant for $\alpha<1$, yielding one chiral topological phase transition, and the topological phases are fixed at $|\mu/t|<2$. 
For arbitrary $\alpha$, $\beta$, although the maximum number of phase transitions remains three, chiral phases are expected, as well as higher values of the topological invariant, bounded by $|\nu|=q$.
In the homogeneous chain, seen in Fig.~\ref{fig: Diagrams}(c), the winding number $|\nu|$ increases up to $q$, such that $|\nu| = 0,1,2$, without chiral phases. 
Fig.~\ref{fig: Diagrams}(d) shows a case setting $\alpha=0$, such that we find chiral phases ($\beta \gtrsim 1.57$) and phases with higher values of the winding number ($\beta \lesssim 1.57$). 
Notice that the phase diagrams of Fig.~\ref{fig: Diagrams} are not symmetric with respect to the exchange $\mu \rightarrow - \mu$, as in the Kitaev chain. 
This $\mu$-reflection is no longer a symmetry, since the gauge transformation $c_j \rightarrow (-1)^j c_j$ exhibits a distinction between even and odd interactions. 
For example, the products  $c_j c_{j+\text{even}}$ yield $c_j c_{j+\text{even}}$ in the even sector, while $c_j c_{j+\text{odd}} \rightarrow - c_j c_{j+\text{odd}}$ in its counterpart, which breaks the mirror symmetry around $\mu$.

In Fig.~\ref{fig: Diagrams}, we separate the critical lines into two different colors (green and orange). 
Green lines represent the quantum phase transitions with central charge $c=1/2$, which are deformed from the nearest-neighbor Kitaev chain critical lines in the presence of next-nearest hoppings. 
We separate in orange the critical points [see Figs.~\ref{fig: Diagrams}(b)-(d)] where the central charge is $c=1$. 
This line separates different topological phases respecting $|\Delta\nu|=|\nu_1 - \nu_2| = 2$. 
Hence, both colors represent topologically inequivalent quantum phase transitions. 
Moreover, notice that in Fig.~\ref{fig: Diagrams}(d) the orange lines cross with the green ones for $\beta \approx 1.57$, generating a critical point with central charge $c=3/2$, which is also not equivalent to the other critical lines.

\section{Majorana formalism} 
\label{sec: Maj-form}

Consider a general quadratic Hamiltonian in 1D, with $L$ sites.
It is written on the $2L$-dimensional Nambu spinor $\boldsymbol{\Psi}^\dagger = (c_1^\dagger \, , \, c_2^\dagger \, \ldots \, c_L^\dagger \, , \, c_1 \, , \, c_2 \, \ldots \, c_L)$ in the Bogoliubov-de Gennes (BdG) space that generates the Hamiltonian $H = \boldsymbol{\Psi}^\dagger \mathbf{h} \, \boldsymbol{\Psi}$, where $\mathbf{h}$ is a $2L \times 2L$ matrix with normal and anomalous terms. 
One strategy for dealing with topological superconductors is to decouple the quadratic Hamiltonian using Majorana operators $b_{2j-1}$ and $b_{2j}$~\cite{Alicea2011Non-AbelianNetworks, Valkov2022TopologicalSystems, Surace2022FermionicApproaches}. 
The basis transformation between the lattice field operators $\widehat{\psi}_j \equiv (c_j \ , \ c_j^\dagger )^T$ and the Majorana lattice spinor $\widehat{b}_j = (b_{2j-1} \ , \ b_{2j})^T$ is expressed as 

\begin{align}
    \widehat{b}_j &= 
    \left( 
    \begin{array}{c}
        b_{2j-1} \\
        b_{2j} 
    \end{array} 
    \right) = \frac{1}{\sqrt{2}} 
    \left( 
    \begin{array}{cc}
        1 & 1 \\
        -i & i
    \end{array} \right)
    \left( 
    \begin{array}{c}
        c_j \\
        c_j^\dagger
    \end{array} \right) = U_j \widehat{\psi}_j, \label{eq: transf-c-eta}
\end{align}
where $U_j$ is a unitary operator. 
Such Majorana operators are generators of a Clifford algebra, given by the anti-commutation relations $\left\{ b_i, b_j \right\} = 2\delta_{ij}$. 
It is convenient to introduce the Majorana spinor of the whole system as $\boldsymbol{b} = (b_1, b_3, \ldots, b_{2L-1}, b_2, b_4, \ldots, \ b_{2L} )^T$, a rotation of the Nambu spinor $\boldsymbol{b} = U \boldsymbol{\Psi} = \oplus_j U_j \widehat{\psi}_j$; and then analyze the properties of the system using this basis. 
The Hamiltonians are expressed on this new basis as $2L \times 2L$ Majorana hopping matrices 

\begin{align}
    H &= i \sum_{j, l = 1}^{2L} b_j \mathcal{M} b_l = \frac{i}{2} \boldsymbol{b}^T \left(  
    \begin{array}{cc}
        0 & M \\
        -M^{T} & 0
    \end{array} \right)
    \boldsymbol{b}. \label{eq: Ham_mat}
\end{align}
The second equality in Eq.~(\ref{eq: Ham_mat}) is only valid for systems in the BDI class [see Sec.~\ref{sec: ERKC}, with the Hamiltonian defined in Eq.~(\ref{eq: Genham KC})] written with real matrix elements; thus, it is an antisymmetric (skew) matrix. 
Consequently, our analysis of the Majorana basis is reduced to the off-diagonal sub-matrices $M$, and the eigenvalues and eigenvectors are obtained using singular value decomposition. 
Thus $M_{j,l} = \sum_{n, m=1}^L (\phi_{nj})^T \Lambda_{nm} \psi_{ml}$, where $j, l = 1, \cdots L$ and $\Lambda_{nm} = \delta_{nm} \xi_n$ are the single-particle eigenenergies matrix, while $\phi_{nl}$ and $\psi_{nl}$ are generalized eigenvectors. 
Hence, the Bogoliubov excitations $\widehat{\gamma}_n = (\gamma_n \ , \ \gamma_n^\dagger)^T$ are determined from Eq.~(\ref{eq: Ham_mat}) using the unitary transformation defined in Eq.~(\ref{eq: transf-c-eta}). 
The excitations are expressed by

\begin{align}
    \widehat{\gamma}_n &= \sum_j U_j 
    \left( 
    \begin{array}{c}
        \phi_{nj} b_{2j-1} \\
        \psi_{nj} b_{2j}
    \end{array}
    \right)
    = U_n \left( 
    \begin{array}{c}
         \tilde{b}_{2n-1} \\
         \tilde{b}_{2n}
    \end{array}
    \right), 
    \label{eq: Bogo-excitations}
\end{align}
where $\phi_{nj}$ and $\psi_{nj}$ are generalized eigenfunctions of $M$~\cite{Lieb1961TwoChain}. 
The operators $\tilde{b}_j$ are called Majorana modes~\cite{Valkov2019Ground-StateNanowire, Valkov2022TopologicalSystems}, and $j = 1, \ldots 2L$ refers to the eigenenergy (in pairs). 
The unpaired Majorana modes~\cite{Kitaev2001UnpairedWires} are described by degenerate single-particle excitations, localized at the edges of the chain. 
This unpairing holds for infinite chains or fine-tuning of parameters, resulting in Majorana zero modes (MZM)~\cite{Alicea2011Non-AbelianNetworks}. 
Finite systems, instead, generally lead to a finite overlap of Majorana modes in the bulk of the chain. 
As a consequence, the Majorana modes bind, realizing a non-local Majorana bound state (non-local Bogoliubov excitation).

\subsection{Ground state fermion parity, effective model, and effective parity}

Of interest to us, the ground state fermion parity $\mathcal{P}$~\cite{Kitaev2001UnpairedWires, DeGottardi2013MajoranaLandscapes, Grabsch2019PfaffianDetection, Roy2026QuantumNanowires} can be determined using the Hamiltonian matrix $M$ defined in Eq.~(\ref{eq: Ham_mat}), 
by its Pfaffian 

\begin{align}
    \mathcal{P} &\equiv \text{sign} \big[ \text{Pf} \big( iH \big) \big] = (-1)^{L(L-1)/2} \text{sign} \big[ \det(M) \big]. \label{eq: parity, def}
\end{align}
The fermion parity changes take place at the degeneracies, with the edge mode eigenenergies being exactly zero. 
We analyze the behavior of the edge modes in finite chains using an effective model that  considers only the relevant topological Majorana bound states.
The effective model is written as 

\begin{align}
    \mathcal{H}_{\text{eff}} &= i \sum_{m=1}^\nu J_m \tilde{b}_{2m-1} \tilde{b}_{2m}, \label{eq: effmodel}
\end{align}
for a negative effective tunneling amplitude $J_m<0$ between Majorana modes across the chain, and $\phi_{mj}$ and $\psi_{mj}$ are coefficients defined in Eq.~(\ref{eq: Bogo-excitations}), where $\nu$ is the number of edge states with nearly-zero energies, i.e., the winding number of Eq.~(\ref{topo_invariant}). 
This effective model describes the presence of a non-local excitation $\gamma_1^\dagger = (\tilde{b}_1 + i \tilde{b}_2)/\sqrt{2}$ in the ground state (also called a Majorana bound state). 
In this effective model, we define the effective parity by considering the $\nu$ Majorana bound states at the edges of the chain by considering individual products, such that 

\begin{align}
    \mathcal{P}_{\text{eff}} = \prod_{m=1}^\nu \mathcal{P}_{\text{eff}(m)} = \prod_{m=1}^\nu \text{sign} (i \tilde{b}_{2m-1} \tilde{b}_{2m}) = \prod_m \text{sign} \Big( \sum_j \phi_{mj} \psi_{mj} \Big). \label{eq: gen-eff-par}
\end{align}
One must consider the Majorana modes product carefully for each Majorana bound state, in order to compute the correct individual effective parity (occupation number) $\mathcal{P}_{\text{eff}(m)}$, since each solution is non-degenerate for finite chains. 

In the presence of one zero-energy mode ($\nu=1$, see App.~\ref{ap: trunc-q}), the Majorana modes are labeled by $\tilde{b}_1$ and $\tilde{b}_2$; thus, the effective parity is given by the sign of the overlap 

\begin{align}
    \mathcal{P}_{\text{eff}} &= \text{sign} \Big( \sum_j \phi_{1j} \psi_{1j} \Big). \label{eq:pareff}
\end{align}

\subsection{Majorana modes moments}

Using the Majorana modes introduced in Eq.~(\ref{eq: Bogo-excitations}), one can also compute the local density of states (LDOS) $\rho_{l}(E)$~\cite{Balatsky2006Impurity-inducedSuperconductors, Fischer2007ScanningSuperconductors, Beenakker2013SearchSuperconductors, Wang2014FateArray, Wakatsuki2014FermionSuperconductor} and the Majorana polarization [$\mathcal{M}_{P;nj} = (\phi_{nj})^2 - (\psi_{n,j})^2$] as diagnostics to verify the topological behavior in finite chains. 
The LDOS represents the density of available states for a given energy $E$ at the $l$-th site, and the Majorana polarization the imbalance between electron- and hole-like constituents per lattice site~\cite{Sticlet2012SpinWires, Sedlmayr2015VisualizingPolarization, Bena2017TestingPolarization}. 
For experiments, one is interested in the particle-like (or hole-like) contributions, which are written as

\begin{align}
    \rho_l(E) &= \sum_{\{ \pm \}} \sum_n |\phi_{nl} \pm \psi_{nl}|^2 \frac{\partial n_F(E \pm \xi_n)}{\partial E}. \label{eq: LDOS}
\end{align}
The LDOS was computed for extensions of the Kitaev chain with multiple bands~\cite{Beenakker2013SearchSuperconductors, Wang2014FateArray, Wakatsuki2014FermionSuperconductor}, and the Majorana polarization was computed for spinful systems. 
Note that for large chains, $|\phi_{nl} - \psi_{nl}| \approx |\phi_{nl}|$ (or $|\psi_{nl}|$) are localized at the boundaries of the chain for $E=0$. 
Consequently, a key argument for analyzing the persistence of topological states in finite systems is the penetration of Majorana edge modes within the chain. 
Based on the spatial analysis of the edge states~\cite{Alecce2017ExtendedPairing, Jager2020EdgeStudy, Ghuneim2026ZeroChain, Pathak2026AnalyticalInward} and the LDOS, we introduce the Majorana moments of the edge modes, defined as 

\begin{align}
    \bar{\phi}_m^{(i)} &\equiv \sum_{j=1}^L j^i (\phi_{mj})^2 = \sum_{j=1}^L (L + 1 - j)^i (\psi_{mj})^2, \label{eq: posmed}
\end{align}
for $i=1,2, \ldots$ 
These quantities give a direct measure to analyze the localization of the edge states, simplifying localization estimates by looking only at the average profile of the Majorana modes in finite systems. 
This furnishes a constraint on the boundary region for the experimental verification of MZM and Majorana bound states in engineered experimental setups, along with numerical computations such as the LDOS in more complicated models, which are very helpful for STM experiments~\cite{Balatsky2006Impurity-inducedSuperconductors, Fischer2007ScanningSuperconductors}. 
In the following, we focus on the first moment ($i=1$), which we call the Majorana average position (MAP). 
It provides the average spatial distribution of the $m$-th Majorana mode, counted from the beginning of the chain. 
As defined in Eq.~(\ref{eq: posmed}), it exhibits values between one and $L$. 
Consequently, the closer $\bar{\phi}_m$ to the boundaries (1 or $L$), the more localized the Majorana modes become.

\section{Majorana bound state properties in finite chains}
\label{sec: finiteC}

We now focus on the behavior of the edge modes in finite chains with open boundary conditions (OBC) over the distinct topological phases shown above. 
We set the chain size to $L = 20$ sites for better visualization. 
We show that the MAP naturally diagnoses the localization of edge states, separating topologically different phases and phase transitions; provided that we know the topology exists. It can also support different localization effects, once the energy splitting is sufficiently large and the single-particle edge state spectral properties decouple energetically (e.g., $\xi_2/\xi_1 \gtrsim 5$).

\subsection{Finite Kitaev chain}

\begin{figure}[!hbt]
    \centering
    \includegraphics[width=0.55\linewidth]{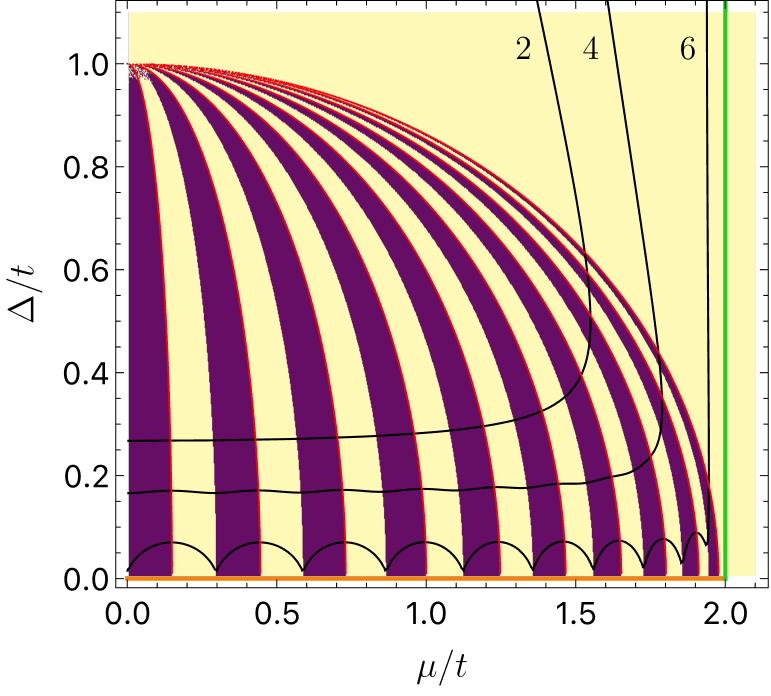}
    \caption{\justifying 
    Truncated phase diagram within the circle of oscillations of the Kitaev chain, showing the effective parity regions where $\mathcal{P}_{\text{eff}}=+1$ (yellow regions) and $\mathcal{P}_{\text{eff}}=-1$ (violet regions) as a function of the parameters $(\mu/t,\Delta/t)$ for a finite chain with $L=20$ sites. 
    Red lines indicate ground-state fermion parity $\mathcal{P}$ switches. 
    Black lines indicate isolines of the MAP from the edge ($\bar{\phi}=2,4$, and 6). 
    Orange line at $\Delta/t=0$ represents the $c=1$ quantum phase transition of the tight-binding model, while the green line at $|\mu/t|=2$ represents the $c=1/2$ Ising transition line. 
    }
    \label{fig: KC(N=20)parity}
\end{figure}

Consider first the original Kitaev finite chain with nearest-neighbor couplings ($q=1$). 
In this case, ground-state fermion parity switches are expected in systems with OBC~\cite{DeGottardi2011TopologicalSystem, Hegde2015QuenchWires, Hegde2016MajoranaChains}. 
The diagonalization procedure of the OBC Kitaev chain follows Sec.~\ref{sec: Maj-form}, and we observe that the spatial profiles of the Majorana modes satisfy the relation $\phi_i = (-1)^{(\mathcal{P}+1)/2} \psi_{L+1-i}$, where the ground-state fermion parity $\mathcal{P}$ determines the sign of the phase of the Bogoliubov excitations defined in Eq.~(\ref{eq: Bogo-excitations}). 
All quantities of interest are depicted joint in Fig.~\ref{fig: KC(N=20)parity}. 
The ground-state fermion parity switches are seen as red curves, which show even- and odd parity sectors within the ellipse $(\Delta/t)^2 + (\mu/2t)^2 = 1$.
This region is often referred to as the \textit{circle of oscillations} in the context of the XY chain~\cite{Lieb1961TwoChain, Barouch1970StatisticalI, Hegde2015QuenchWires, Hegde2016MajoranaChains, Perez2017PolynomialWires}. 
Within it, a clear pattern of effective parity becomes evident. 
Observe that at the places where the ground-state fermion parity switches (red lines), it forces the effective parity to also switch since the energy degeneracy of Majorana modes corresponds to an occupied Bogoliubov excitation becoming unoccupied; we call this an \textit{energetic switch}. 
In addition, the effective parity switches once more in the middle of two adjacent ground-state fermion parity switches (from left to right: yellow to violet areas without a red line). 
This extra effective parity change results from modifications of the Majorana modes overlap, which we call it a \textit{functional switch}. 
Therefore, we have separated these transitions into two distinct origins. 

Beyond parity, the MAP, represented by the black isolines in Fig.~\ref{fig: KC(N=20)parity}, illustrates how the parameters determine the behavior of the topological edge states. 
It behaves as expected along the topological domain (including within the circle of oscillations), except for $\Delta/t \approx 0$; see bottom area of Fig.~\ref{fig: KC(N=20)parity}.
If $\Delta/t = 0$, then it corresponds to a tight-binding chain (or equivalently, the XX spin model~\cite{Jin2004QuantumConjecture}), where the MAP exhibits maxima and minima at the effective parity switches. 
It has a maximum value when there is an \textit{energetic switch}, while its minimum value occurs near a \textit{functional switch}. 
This behavior extends to $\Delta/t\neq0$, and consequently the edge modes are, on average, more localized when the ground state parity is fixed rather than close to a parity change.

\subsection{Next-nearest neighbor limit}

\begin{figure}[!hbt]
    \centering
    \begin{minipage}{0.65\linewidth}
        \centering
        \includegraphics[width=\linewidth]{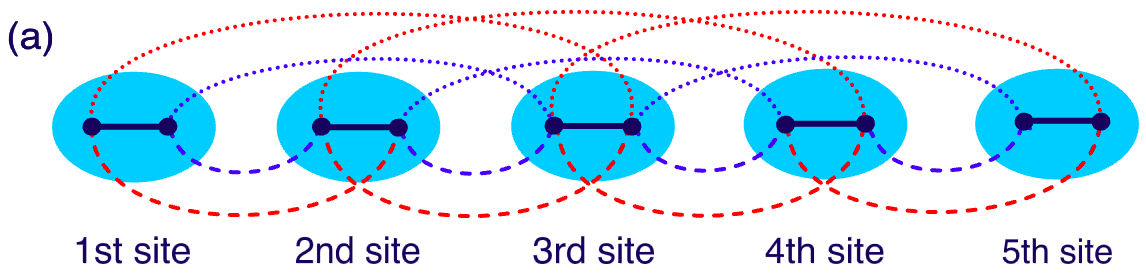}
    \end{minipage} 
    \begin{minipage}{0.75\linewidth}
        \centering
        \includegraphics[width=\linewidth]{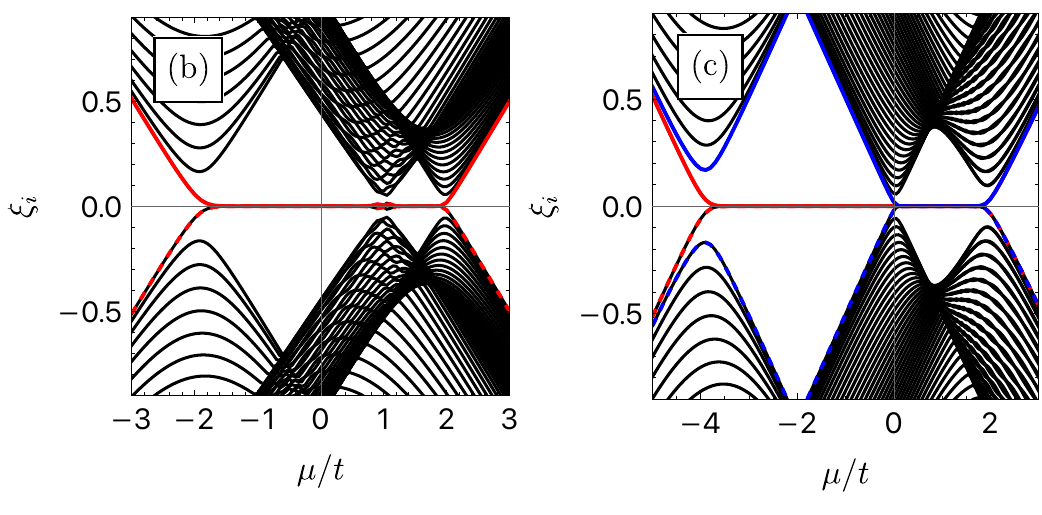}
    \end{minipage}
    \caption{\justifying (a) 
    Schematic example of a finite open chain with $L=5$ sites in the trivial phase of the next-nearest neighbor ($q=2$) model. 
    Nearest-neighbor couplings are represented by red and blue dashed lines, while next-nearest ones, by red and blue dotted lines. The blue solid lines represent the intrasite chemical potential coupling. (b) and (c) 
    Eigenenergies of the $q=2$ Kitaev chain with $L=20$ sites as a function of $\mu/t$ are seen in the limits (b) $\alpha=0$ and $\beta \rightarrow \infty$, and (c) the homogeneous case $\alpha=\beta=0$, where red and blue curves represent the two lowest eigenmodes.}
    \label{fig: scheme2modes_NNenergies}
\end{figure}

Taking the phase diagrams in Fig.~\ref{fig: Diagrams}, we are now interested in characterizing the zero-energy modes for finite chains with next-nearest neighbor interactions, as depicted in Fig.~\ref{fig: scheme2modes_NNenergies}(a). 
Notice in Fig.~\ref{fig: scheme2modes_NNenergies}(b) the existence of one zero energy mode in OBC, while in Fig.~\ref{fig: scheme2modes_NNenergies}(c) there are two zero energy modes in the interval $0<\mu/t<2$.
Hence, we separate two distinct limits of the phase diagrams obtained in Fig.~\ref{fig: Diagrams}: Case (\textit{1.}) stands for the phase diagram of Fig.~\ref{fig: Diagrams}(b) with chiral phases; Case (\textit{2.}) refers to quantum phase transitions with higher values of the winding number ($|\nu|\geq 2$), namely, for the model of next-nearest neighbors, as in  Figs.~\ref{fig: Diagrams}(c) and 1(d). 

\subsubsection{Case(1): Chiral Majorana modes}

\begin{figure}[!thb]  
    \centering
    \begin{minipage}{0.475\linewidth}
        \includegraphics[width=\linewidth]{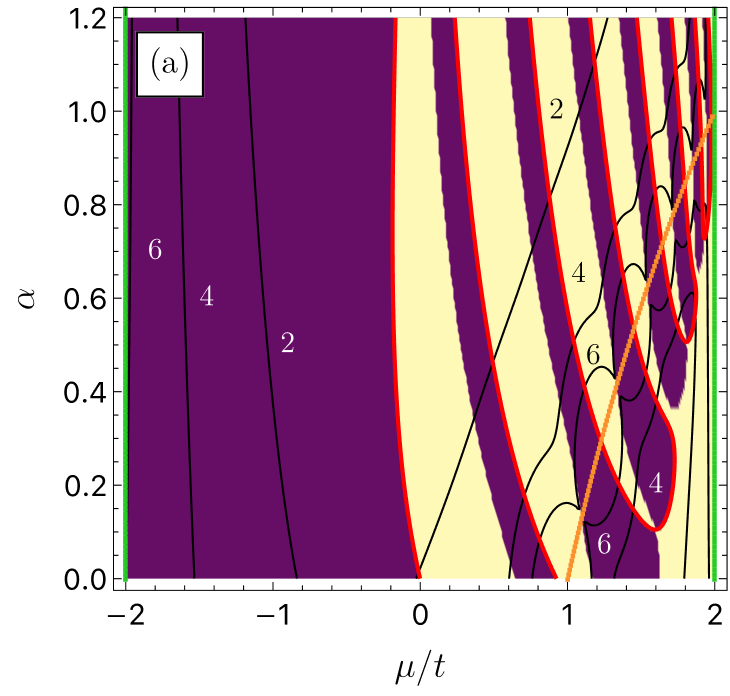}
    \end{minipage}
    \begin{minipage}{0.45\linewidth}
        \centering
        \includegraphics[width=\linewidth]{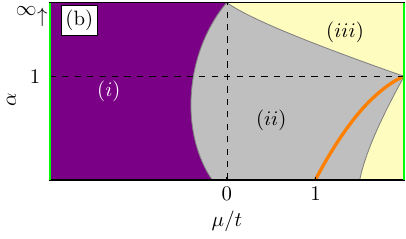}
    \end{minipage}
    \caption{\justifying 
    (a) Color diagram for a $q=2$ chain with $L=20$ sites, exhibiting the values of $\mathcal{P}_{\rm eff}=-1$ (violet regions) and $\mathcal{P}_{\rm eff}=+1$ (yellow regions) in the chiral limit $\beta \rightarrow \infty$ of Fig.~\ref{fig: Diagrams}(b). 
    Green lines display the quantum phase transitions observed in the Kitaev chain, while the orange line shows transitions from $\nu=-1$ to $\nu=+1$. 
    Black solid isolines indicate the MAP of the modes from the border of the chain ($\bar{\phi} = 2,4,6$). 
    Red curves exhibit ground state fermion parity switches ${\mathcal P}=0$. 
    (b) Schematic diagram of the effective parity for a generic case with any number of sites. Violet region $(i)$ and yellow regions $(iii)$ with same color schemes as in part (a). A new gray color delimits a region $(ii)$, where ground-state and effective parities change.}
    \label{fig: exact-effparb->infty}
\end{figure}

The topological phase diagram in the limit $\beta \rightarrow \infty$, shown in Fig.~\ref{fig: Diagrams}(b), exhibits two  (chiral) topological phases, distinguished by opposite signs\footnote{In the Majorana basis framework, there is no simple combination of parameters to represent unpaired zero-energy modes at the edges of the chain (contrary to the Kitaev chain, by selecting $\mu/t = 0$ and $\Delta/t =1$).}. We plot this limit again in Fig.~\ref{fig: exact-effparb->infty} adding further informational details. 
Notice that the phases are separated by revivals of the edge modes and has exactly zero energy at the critical point $\mu/t=0$; differently from standard quantum phase transitions, which  close the gap for infinite systems only, this topological phase transition occurs also for finite systems. 

The effective parity $\mathcal{P}_{\text{eff}}$, the ground state fermion parity ${\mathcal P}$, and the MAP $\bar\phi$ are shown in Fig.~\ref{fig: exact-effparb->infty}(a). 
A similar pattern is observed here with additional effective parity switches between ground state fermion parity ones, as in Fig.~\ref{fig: KC(N=20)parity}. 
In addition, close to the quantum phase transition (orange line), the changes in effective parity mark local extreme values in the MAP, along with its maximum localization at \textit{functional switches} and its minimum localization at \textit{energetic switches}.
A schematic effective parity diagram is also included in Fig.~\ref{fig: exact-effparb->infty}(b) for a general number of sites and shows three different regions. 
The areas $(i)$ and $(iii)$ exhibit the usual parity sectors, in the sense that they are also present in the Kitaev chain limit $(\alpha \rightarrow \infty)$ (see Fig.~\ref{fig: KC(N=20)parity}, $\Delta/t = 1$). 
Notice that areas $(i)$ and $(iii)$ have a fixed effective parity: $(i)$ the left side of the plot (violet region) has $\mathcal{P}_{\text{eff}}=-1$; $(iii)$ the right parts of the plot (yellow regions) have $\mathcal{P}_{\text{eff}}=+1$. 
The remaining gray area $(ii)$ exhibits many ground state fermion parity switches and effective parity switches. 

Furthermore, the two topological phases are distinguished by the configuration of the Majorana modes. 
When passing from one phase to the other, the localized Majorana modes swap, i.e., $(\phi_1, \psi_1, \nu = -1) \rightarrow (\psi_1, \phi_1, \nu = +1)$ they switch roles, indicating chiral behavior. 
In each of the chiral phases, $\nu=-1$ or $+1$, a preference is observed for an effective parity: for $\nu=-1$, the effective parity is mostly even, while for $\nu = +1$, the effective parity is preferably odd. 
The orange line with central charge $c=1$ vanishes at $\mu/t = 2$ and $\alpha=1$. 
Besides, the green and orange lines have $\omega =-1$, and none of them host MZM, as seen in Fig.~\ref{fig: exact-effparb->infty}(a), where the MAP indicates delocalization, and Fig.~\ref{fig: scheme2modes_NNenergies}(b), where the Majorana mode energy becomes massive close to the $\mu/t=1$. 
Additionally, we notice that fluctuations of the ground state parity are expected along such phase transition for arbitrary finite systems, see Fig.~\ref{fig: exact-effparb->infty}(b) and Fig.~\ref{fig: KC(N=20)parity}.

\subsubsection{Case (2): Two-Majorana-mode phase}
\label{sub: 2Mm}

Consider now $\alpha=\beta$, of the phase diagram shown in Fig.~\ref{fig: Diagrams}(c) and the eigenenergies with OBC seen in Fig.~\ref{fig: scheme2modes_NNenergies}(c), with $\alpha=\beta=0$. These figures show a difference in the behavior of each topological phase ($\nu = -1$ and $\nu = -2$), which admits one or two Majorana bound states, with energies $\xi_1, \xi_2$ (with Majorana modes $\phi_1$, $\psi_1$ and $\phi_2$, $\psi_2$), respectively.
This two-Majorana modes phase is also qualitatively distinct from the one observed in chiral phases, described previously, in Figs.~\ref{fig: scheme2modes_NNenergies}(b) and \ref{fig: scheme2modes_NNenergies}(c). 

Figure~\ref{fig: pos-par-ab,b0} shows the properties of single-particle excitations in the two-Majorana modes phase for a finite system ($L=20$ sites) and for $\alpha=\beta$. 
The MAP of each mode is also well-behaved when the effective parity is fixed, see Fig.~\ref{fig: pos-par-ab,b0}(a) and \ref{fig: pos-par-ab,b0}(b). 
In this regime, the energies of the Majorana modes are well separated, and both Majorana modes do not interact. 
When the energies are close to each other, the effective parity fluctuates, and the MAP of both modes is closer, exhibiting kinks at the points where both modes have the same MAP in the parameter space. 
This translates in a larger overlap and interference between the edge modes. 
Moreover, the effective parity fluctuates in this region, see Fig.~\ref{fig: pos-par-ab,b0}(c). 
Additionally, we note that in the two-Majorana modes phase, the MAP indicates the existence of an SPT Majorana mode confined at the edge of the system crossing the phase transition with central charge $c=1/2$ (green line). 

\begin{center}
    \begin{figure}[!htb]
        \begin{minipage}{0.675\linewidth}
        \begin{minipage}{\linewidth}
            \centering
            \includegraphics[width=1\linewidth]{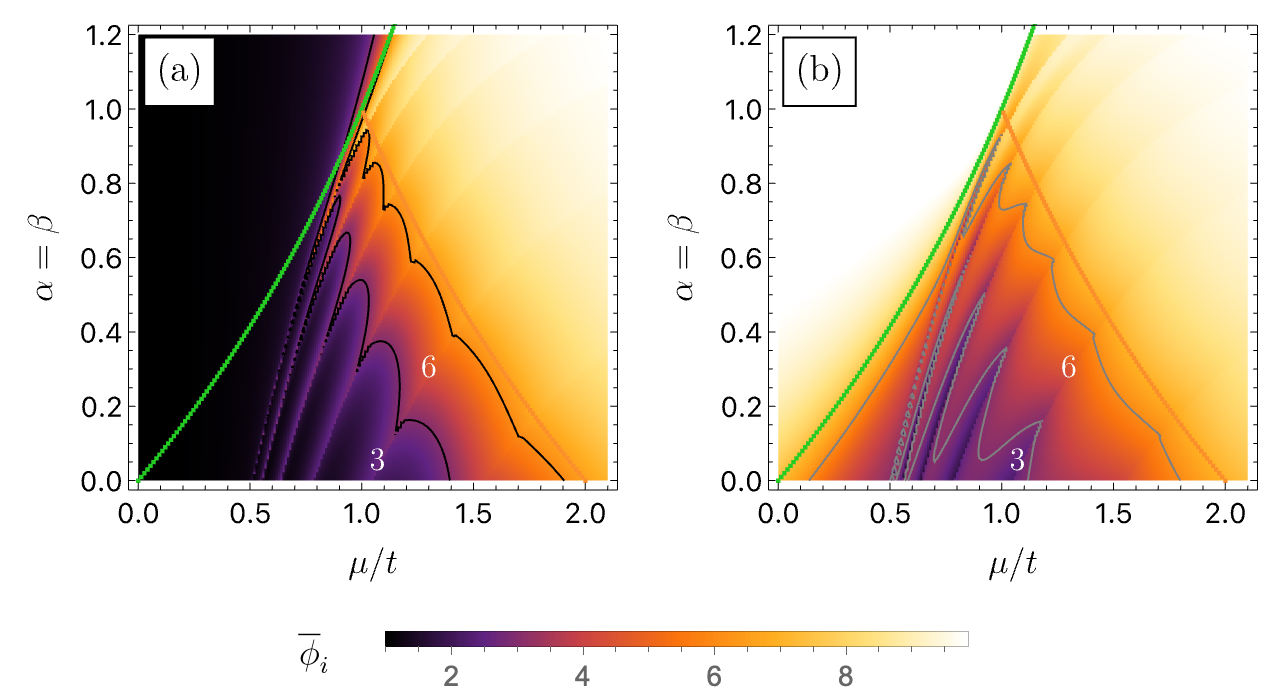}
        \end{minipage}
        \end{minipage}
        \begin{minipage}{0.32\linewidth}
            \centering
            \includegraphics[width=\linewidth]{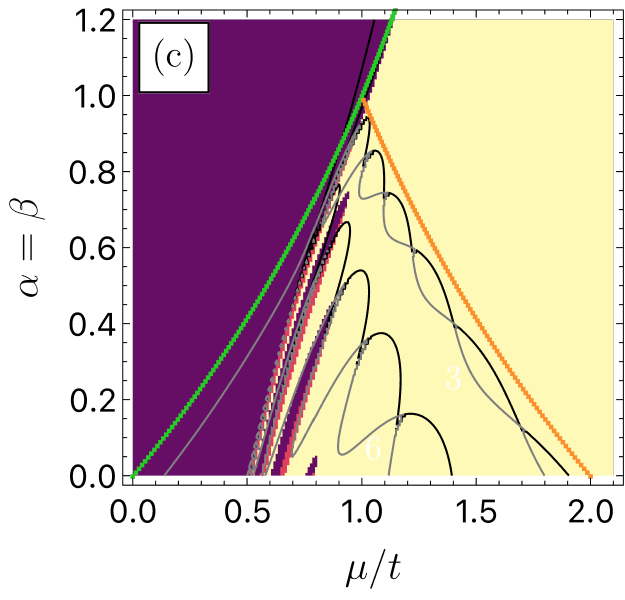}
            \vspace{0.225cm}
        \end{minipage}
        \center
        \caption{\justifying MAPs
        $\bar{\phi}_i$ from the border of the chain and the effective parity $\mathcal{P}_{\text{eff}}$ for chains with $L = 20$ sites and for $\alpha=\beta$: Panels 
        (a) and (b) represent the structure of each Majorana mode $\bar{\phi}_1$ and $\bar{\phi}_2$, respectively.
        Black (gray) solid isolines represent  $\bar{\phi}_1, (\bar{\phi}_2) = 3, 6$.
        Panel (c) exhibits the effective parity changes $\mathcal{P}_\text{eff} = \mathcal{P}_{\text{eff}(1)} \mathcal{P}_{\text{eff}(2)}$, emphasizing individual parity changes (violet, brown, and yellow).  
        Green lines portray the $c=1/2$ quantum phase transitions, while the orange ones are due to the inclusion of next-nearest neighbors hopping and pairing, that gives $c=1$. 
        In (c), the violet color represents the parity sector with   $\mathcal{P}_{\text{eff}(1)} = +1$ and $\mathcal{P}_{\text{eff}(2)} = -1$, the brown color represents the sector $\mathcal{P}_{\text{eff}(1)} = -1, \mathcal{P}_{\text{eff}(2)} = +1$, while the yellow color represents the even parity sectors $\mathcal{P}_{\text{eff}(1)} = \mathcal{P}_{\text{eff}(2)} = \pm 1$.}
        \label{fig: pos-par-ab,b0}
    \end{figure}
\end{center}

\begin{center}
    \begin{figure}[!htb]
        \begin{minipage}{\linewidth}
            \centering
            \includegraphics[width=0.95\linewidth]{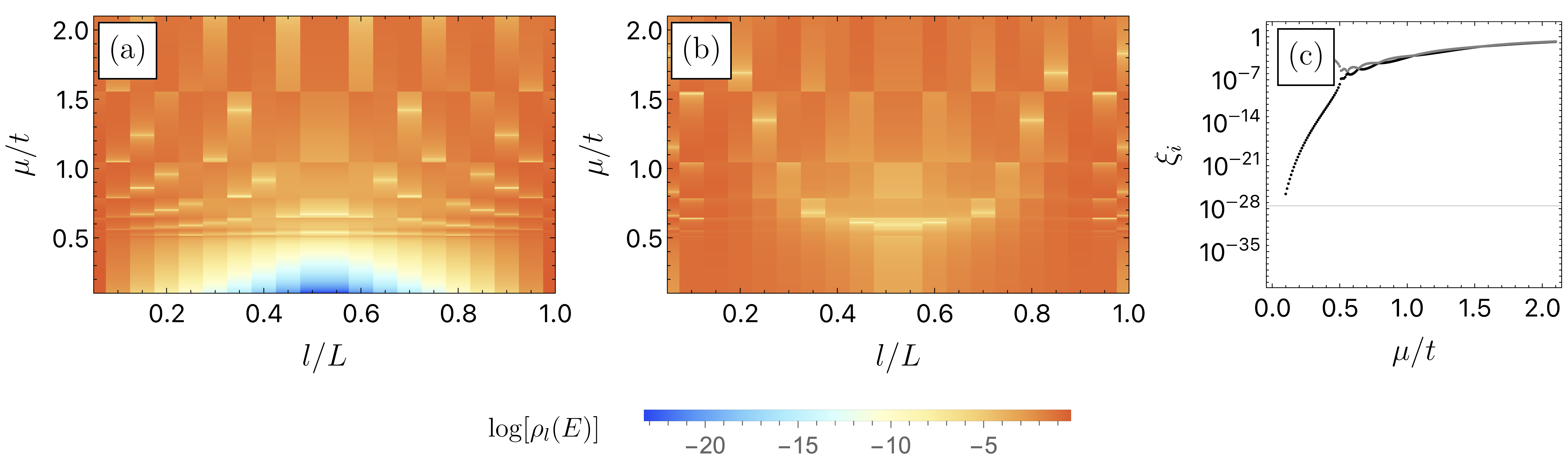}
            \includegraphics[width=0.95\linewidth]{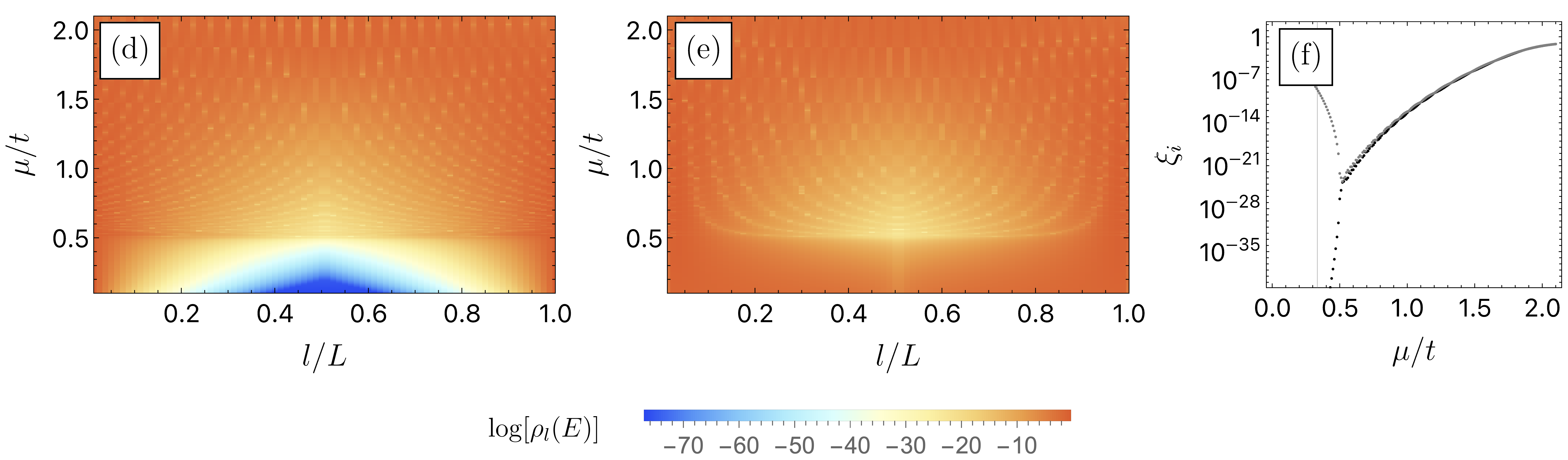}
            \includegraphics[width=0.95\linewidth]{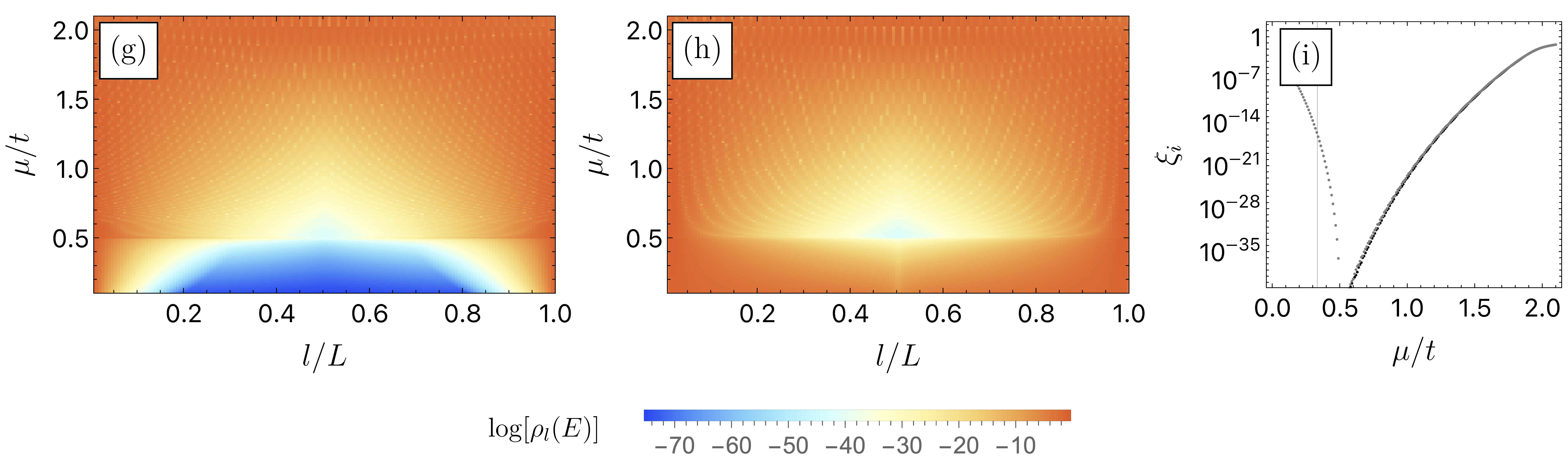}
        \end{minipage}
        \center
        \caption{\justifying 
        Normalized LDOS along the chain (colored panels) and energy of the two Majorana bound states as a function of the chemical potential (rightmost column). 
        Horizontal panels: (a)-(c) $L=20$ sites, (d)-(f) $L=80$ sites, and (g)-(i) $L=150$ sites; vertical panels:
        (a), (d) and (g) show the logarithm of the LDOS ($\rho_j(E)$) setting the energy at $E = \xi_1$; panels
        (b), (e), and (h) the same at energy  $E=\xi_2$; panels
        (c), (f), and (i) show the energy of the edge Majorana bound states as a function of the chemical potential. 
        Note the tendency $|\xi_1 - \xi_2| \rightarrow 0$ close to $\mu/t \sim 1/2$.
        }
        \label{fig: LDOS-ab0}
    \end{figure}
\end{center}

Finally, in Fig.~\ref{fig: LDOS-ab0}, the LDOS is computed for the next-nearest Kitaev chain in the two-Majorana modes phase ($\nu = -2$) seen in Fig.~\ref{fig: Diagrams}(c), considering the energies of the edge modes $E \approx \xi_1$ and $E \approx \xi_2$. 
We observe that the energies of the excitations and the spatial distribution of the Majorana modes are not equal in a finite range of the chemical potential [$0 < \mu/t \lesssim 1/2$, see Fig.~\ref{fig: LDOS-ab0}(c), (f), and (i) and Fig.~\ref{fig: pos-par-ab,b0}(a) and (b)]. 
As the chemical potential increases towards $\mu/t \sim 1/2$, the energy of one Majorana bound state ($\xi_1$) increases while the other decreases ($\xi_2$), getting closer to each other $|\xi_1 - \xi_2| \rightarrow 0$. 
In this region, the MAP correctly captures the real-space properties observed when computing the LDOS, and the edge state physics is described with only one Majorana mode localized at the boundary\footnote{The second Majorana edge state is spread over the system, see the bottom of Figs.~\ref{fig: LDOS-ab0}(b), (e), and (h).}. 

Approximately at $\mu/t \approx 1/2$, the eigenenergies become degenerate, and for larger values of the chemical potential ($1/2 \lesssim \mu < 2$), the energies remain close to each other ($\xi_1 \sim \xi_2$), converging towards zero as the size of the system increases. 
In this region, the energies of the modes are of the same order $(\xi_2/\xi_1 \approx 1)$, and despite the MAP of the individual modes remaining localized and well-behaved (see Fig.~\ref{fig: pos-par-ab,b0}), they cloak mutual interference, which is only seen when analyzing the LDOS (see Figs.~\ref{fig: LDOS-ab0}(a)-(b), (d)-(e), and (g)-(h)). 
Using the LDOS, we observe a spreading of the real-space distribution across the entire chain, which annihilates the localized character of the topological phases in the two-Majorana modes,  see Fig.~\ref{fig: LDOS-ab0}(d) and (e) at the first degenerate energy. 
Figure~\ref{fig: LDOS-ab0} indicates that the two Majorana modes are confined at the boundaries when considering chains in thermodynamic limit, as seen in Fig.~\ref{fig: LDOS-ab0}(g) and (h), which in the range of $1/2<\mu/t\lesssim 1$ the edge modes remain localized in the first few sites of the chain ($l/L<0.3$). 
In this limit, the first third of the chain ($j<L/3$) has $\rho_j(E)\approx 10^{-2}$, and in the bulk ($L/3<j<2L/3$) has $\rho_{j}(E) \approx 10^{-30}$. 
Instead, for $\mu/t\gtrsim 1$, the modes spread all over along the chain. 

Hence, engineering small chains with next-nearest neighbors interactions culminates in an  interference between modes and delocalization of the edge states (for $\mu/t>1/2$), or the existence of only one Majorana mode (for $\mu/t < 1/2$).

\section{Conclusions}

In this work, we provided an analytical method for constructing phase diagrams, based on discontinuities of the topological angle, for 1D
extensions of Kitaev superconducting chains belonging to the BDI symmetry class with extended-range interactions. 
We derived the main properties of their phase diagrams, including the number of phases and the maximum value of the topological index using Chebyshev polynomials. 

We focused on the next-nearest-neighbor scenario; where we explored unusual topological phases obtained in the phase diagrams. 
Two distinct limits were of great interest: the chiral phases, with the topological index changing sign ($\nu=\pm1$); and the multiple-Majorana-mode phases, where the topological index takes a value greater than one (in our case, $|\nu| \leq 2$). 
We also highlight the existence of multiple topologically distinct quantum phase transitions. 

We also proposed a physical quantity, based on Majorana polarization and the LDOS to analyze the real-space distribution of the localized Majorana bound states in finite open systems. 
We numerically diagonalized the Hamiltonian of finite open chains to obtain the Majorana bound states and their respective Majorana modes in both the Kitaev chain and extended models, obtaining reliable results. 
The key properties of the edge modes and their relationship to the many-body ground state were computed by introducing the MAP (Majorana average position) and the effective parity $\mathcal{P}_{\text{eff}}$, with the latter relating the behavior of the ground state fermion parity switches to the overlap of the Majorana modes.
In the case of two edge modes, such as the two-Majorana modes phase, we extended the effective parity to consider all $q$ edge modes. 

Our numerical calculations revealed a relationship between the MAP, the effective parity, and the ground state fermion parity for finite chains in the Kitaev chain and the chiral next-nearest neighbors Kitaev chain. 
Specifically, we show that the effective parity changes at the extreme points of the MAP (local maxima and minima), with the ground state fermion parity changing at the maxima of the MAP. 
We further claim that the changes in the effective parity of the topological modes are of two distinct origins, which we call \textit{energetic switches} and \textit{functional switches}. 
Moreover, we observe a topologically inequivalent quantum phase transition with central charge $c=1$. 

In the two-Majorana modes phase, ground state fermion parities are fixed, while effective parity switches are present when the edge states are nearly degenerate. 
In these phases, the degeneracies of the Majorana modes play a significant role, only observed in the LDOS. 
The LDOS reveals the interference, leading to delocalization of the multiple Majorana modes across the chain. 
We observe that the interference of the two Majorana modes dramatically changes the LDOS relative to the MAP. 
Hence, in the non-interacting limit, the existence of multiple Majorana modes at the edge are not described by a single-particle property. 
Additionally, we find in the two-Majorana modes phase a topologically distinct quantum phase transition with $c=1/2$ that hosts a localized Majorana mode, observed using the MAP.

\subsection*{Acknowledgments}

We acknowledge support for this work by the Brazilian research agencies CAPES and CNPq,  and the Gravitation grant agency Delta ITP (Utrecht). 
We also thank many fruitful discussions with Leonardo Prauchner and Profs.\  Angela Foerster, David Möckli, and Sérgio Garcia Magalhães over the production of this work.

\appendix

\vspace{12pt}

\noindent
{\bf Appendix}

\section{Topological Invariant: truncated $q$-range} \label{ap: trunc-q}

We separated the results that provide information on the topological phases according to the properties of the single-particle energy and the topological angle [defined from $\mathbf{n}_k^q = (n_{k,x},n_{k,y}) \rightarrow (\xi_k, \vartheta_k)$]

\begin{align}
    \xi(k) &= \sqrt{n_{k,x}^2 + n_{k,y}^2}\, ; &
    n_{k,x} &= \xi(k) \cos{\vartheta_k}\, ; \\
    \vartheta_k &= \tan^{-1}\frac{n_{k,y}}{n_{k,x}}\, ; &
    n_{k,y} &= \xi(k) \sin{\vartheta_k}\,,
\end{align}
so we look for information on $n_x^q$, $n_y^q$ separately, and also on energy $\xi(k)$. 

Let us first analyze the contribution of $\cos{\vartheta_k}$ to the construction of topological phase diagrams. 
Consider $\cos{\vartheta_k} = 0$ and recall the general definition of Chebyshev polynomials of the first kind \cite{Zwillinger1999CRCFormulae} 

\begin{align}
    \cos{nx} &\equiv T_n(\cos{x}), \nonumber\\
    T_n(x) &= \sum_{j=0}^{ \lfloor \frac{n}{2} \rfloor} \binom{n}{2j} x^{n-2j} \br{x^2 - 1}^j,
\end{align}
for the floor function $\lfloor x \rfloor$. 
We write $\cos{\vartheta_k}$ from the polynomials as

\begin{align}
    &\cos{\vartheta_k} = \frac{n_{k,x}}{\xi_k} = - \frac{1}{\xi_k} \left[ \mu - 2t \sum_{\ell=1}^q \frac{1}{\ell^\beta} T_\ell(\cos{k}) \right] = 0\, ; \label{ap: eq-Discon_gen} \\
    &0 = \frac{\mu}{2t} + \sum_{\ell = 1}^q \frac{1}{\ell^\beta} \sum_{j=0}^{ \lfloor \frac{\ell}{2} \rfloor} \sum_{m=0}^j (-)^{j-m} \binom{\ell}{2j} \binom{j}{m}  x^{\ell-2(j-m)} \, , \nonumber
\end{align}
where $x = \cos{k}$, this turns into a polynomial equation of degree $q$, therefore, we already expect a set of at most $q$ real roots, according to the Fundamental Theorem of Algebra. 
These solutions imply that the quasimomenta $k_i = \arccos\br{x_i}$ are doubled and symmetrical over the BZ. 
Numerically, we find that these real solutions satisfy the condition $|x_i| < 1$.

Consider $n \leq q$ real solutions to the polynomial Eq. (\ref{ap: eq-Discon_gen}). 
We can sort these solutions from the quasimomenta $\{ k_i \}$ (with the maximal and minimal elements $k_{max} = \text{max} \{ k_i \} , k_{min} = \text{min} \{ |k_i| \}$ respectively) and label them as $l_i$

\begin{align}
    -\pi =l_0 \leq l_1 = -k_{max} \leq \ldots \leq l_m = -k_{min} \leq 0; \nonumber \\
    0  \leq l_{m+1} = k_{min} \leq \ldots \leq l_{2m} = k_{max} \leq \pi = l_{2m+1}, \label{ap: def-quasi-li}
\end{align}
with $l_0=-\pi$ and $l_{2m+1}=\pi$ are terms always present.
These solutions do not imply the presence of topological phases. 
Instead, we know that $\cos{\vartheta_k}$ gives us $2m$ discontinuities in the BZ, which allows phases with a topological index of up to $q$. 
The realization of such phases depends on the sign of $\sin{\vartheta_k}$. 
Moreover, these $q$ solutions give us an upper bound on the maximum number of phase transition lines. 

Consider $q=1$ (Kitaev chain), there are two constant Ising lines. 
Now for $q=2$, seen in Fig.\ \ref{fig: Diagrams}(a)-(d), there are at most three phase-transition lines. 
This is naturally extended for the general case with $q$-range interactions, which exhibit $q+1$ phase-transition lines, verifying the results obtained in Refs.~\cite{Jones2019AsymptoticSystems, Jones2023Bulk-BoundaryChains}.

Consider now $\sin{\vartheta_k}=0$ at some point in the BZ. This term reveals changes in the sign of the angle $\vartheta_k$, and it is important for Eq.\ (\ref{eq: disc nu}) to compute the correct value of the winding number and the topological phase transition. 
Since $\sin{nk}$ is related to Chebyshev polynomials of the second kind $U_n(x)$, as defined in \cite{Zwillinger1999CRCFormulae},

\begin{align}
    \sin{nx} &=  U_{n-1}(\cos{x}) \sin{x}\,, \nonumber \\
    U_n(x) &= \sum_{r=0}^{\lfloor \frac{n}{2} \rfloor} (-1)^r \binom{n-r}{r} (2x)^{n-2r}\,.
\end{align}
Assuming again $\cos{k}=x$, we have the following polynomial equation (considering $\Delta \neq 0$)

\begin{align}
    &\sin{\vartheta_k} = \frac{n_{k,y}}{\xi_k} = 2\Delta \sin{k} \sum_{\ell=1}^q \frac{1}{\ell^\alpha} U_{\ell-1}(\cos{k}) = 0\,; \\
    &0 = \sqrt{1-x^2} \sum_{\ell=1}^q \frac{1}{\ell^\alpha} \sum_{r=0}^{\lfloor \frac{\ell-1}{2} \rfloor} (-)^r \binom{\ell-1-r}{r} (2x)^{\ell-1-2r}. \nonumber
\end{align}
Beyond the trivial solution at $\sin{k}=0$, there can be $q-1$ sign changes, which were numerically verified, giving real solutions for $|x_i|<1$. 

Since $\sin{\vartheta_z} = 0$ for some $z=\arccos{x_i}$ in the BZ, the emergence of a non-trivial winding around the origin occurs if there are $l_{i},l_{i+1}$ such that $l_i<z<l_{i+1}$ (there is a change in the sign of $\sin{\vartheta_k}$). 
This result is trivially extended to the case of many zeros of $\sin{\vartheta_k}=0 \rightarrow \{ z_i \}$, provided that multiple roots $\{ l_i \}$ are defined as in Eq. (\ref{ap: def-quasi-li}). 

We know that $\sin{\vartheta_k}$ does not depend on any other parameter than $\alpha$.
So, if for a certain value of $\alpha$, such as  $\vartheta_{l_{i+1},-} - \vartheta_{l_{i},+} = 0$ or $\vartheta_{l_{i+1},-} - \vartheta_{l_{i},+} = \pi$, then the winding number is just

\begin{align}
    \nu &= \frac{1}{2\pi} \br{ \br{ \varphi_{-k_0^-} - \varphi_{-k_\pi^+}} + \br{ \varphi_{k_\pi^-} - \varphi_{k_0^+}  } } = \pm 1, \label{eq: chiral nu}
\end{align}
Now, if there are other discontinuities in the BZ $\{ k_1, \ldots , k_n \}$ for $1 \leq n \leq q$ where $\sin{\varphi_{k_n}^q} = 0$ and

\begin{align*}
    \varphi_{l_i}^q < \varphi_{k_n}^q < \varphi_{k_{n'}}^q < \ldots < \varphi_{l_{i+1}}^q,
\end{align*}
for an odd number of $k_{n}$`s, then $\varphi_{l_{i+1}^-} - \varphi_{l_i^+} = \pi$ and since Eq.\ (\ref{eq: chiral nu}) is valid, $\nu$ is neither $-1$ nor $+1$, it is an integer greater than one.





\baselineskip 4.3mm

\bibliographystyle{iopart-num}

\bibliography{references, fix}

\end{document}